\documentclass[%
reprint,
superscriptaddress,
 amsmath,amssymb,
 aps,
prl
]{revtex4-2}

\usepackage{graphicx}% Include figure files
\usepackage{dcolumn}% Align table columns on decimal point
\usepackage{bm}% bold math
\usepackage{braket}
\usepackage{mathrsfs}  
\usepackage{xfrac}
\renewcommand\bra[1]{{\langle{#1}|}}
\makeatletter
\renewcommand\ket[1]{%
  \@ifnextchar\bra{\k@t{#1}\!}{\k@t{#1}}%
}
\newcommand\k@t[1]{{|{#1}\rangle}}
\makeatother

\begin{document}

\preprint{APS/123-QED}

\title{Quantum sensing in the fractional Fourier domain}% Force line breaks with \\

\author{Swastik Hegde}
\affiliation{Illinois Quantum Information Science and Technology Center (IQUIST), University of Illinois at Urbana-Champaign, Urbana, IL 61801}
\affiliation{Center for Biophysics and Quantitative Biology, University of Illinois at Urbana-Champaign, Urbana, IL 61801}
\author{David J. Durden}
\affiliation{Illinois Quantum Information Science and Technology Center (IQUIST), University of Illinois at Urbana-Champaign, Urbana, IL 61801}
\affiliation{Department of Chemistry, University of Illinois at Urbana-Champaign, Urbana, IL 61801}
\author{Lakshmy Priya Ajayakumar}
\affiliation{Illinois Quantum Information Science and Technology Center (IQUIST), University of Illinois at Urbana-Champaign, Urbana, IL 61801}
\affiliation{Department of Chemistry, University of Illinois at Urbana-Champaign, Urbana, IL 61801}
\author{Rishi Sivakumar}
\affiliation{Department of Physics, University of Illinois at Urbana-Champaign, Urbana, IL 61801}
\author{Mikael P. Backlund}
\email{mikaelb@illinois.edu}
\affiliation{Illinois Quantum Information Science and Technology Center (IQUIST), University of Illinois at Urbana-Champaign, Urbana, IL 61801}
\affiliation{Center for Biophysics and Quantitative Biology, University of Illinois at Urbana-Champaign, Urbana, IL 61801}
\affiliation{Department of Chemistry, University of Illinois at Urbana-Champaign, Urbana, IL 61801}
\date{\today}% It is always \today, today,

\begin{abstract}
Certain quantum sensing protocols rely on qubits that are initialized, coherently driven in the presence of a stimulus to be measured, then read out. Most widely employed pulse sequences used to drive sensing qubits act locally in either the time or frequency domain. We introduce a generalized set of sequences that effect a measurement in any fractional Fourier domain, i.e. along a linear trajectory of arbitrary angle through the time-frequency plane. Using an ensemble of nitrogen-vacancy centers we experimentally demonstrate advantages in sensing signals with time-varying spectra.
\end{abstract}

\maketitle

%\tableofcontents

\section{\label{section_introduction}Introduction\\}
Exciting progress has been made in the deployment of quantum technologies to sense physical phenomena with improved resolution, sensitivity, and/or robustness \cite{Degen_RevModPhys_2017,pirandola2018advances,crawford2021quantum,aslam2023quantum,bass2024quantum}. An important subset of these techniques operate by coherently driving a sensing qubit in the presence of a stimulating (deterministic or stochastic) waveform, then inferring the effect of the stimulus on the coherence \cite{Degen_RevModPhys_2017}. In most implementations to date the qubit is driven periodically, and consequently the measurement of the waveform is localized in the frequency domain. Time-varying spectra abound in nature and technology, however, and so the most pertinent information about a signal might not be neatly extracted in the frequency domain. In this work we adapt concepts from classical time-frequency analysis \cite{CohenTimeFreq} for this context by introducing a set of driving protocols that correspond to measurements in any fractional Fourier domain \cite{OzaktasFRFT}, of which the ordinary frequency domain is a special case.

We consider a sensing qubit subject to a Hamiltonian of the general form
\begin{equation}
    \mathscr{H} = \frac{\hbar}{2}\left( \omega_0 + g(t) \right)\sigma_z + \mathscr{H}_c(t),
\end{equation}
where $\hbar \omega_0$ is some initial energy separation of the two qubit levels, $g(t)$ includes both the stimulus to be sensed and possibly background, and $\sigma_z$ is the Pauli-$z$ operator. For now let us assume that $g(t)$ is deterministic, though we will consider stochastic processes later. Coherent manipulation of the qubit with a carefully selected control Hamiltonian $\mathscr{H}_c(t)$ allows one to tunably couple to a feature of interest while simultaneously decoupling from unwanted background. We let $\mathscr{H}_c(t)$ take the form of a driving field polarized perpendicular to $z$, at resonance frequency $\omega_0$, and with time-dependent amplitude proportional to $\Omega(t)$. After invoking the rotating-wave approximation and applying the appropriate unitary transformations \cite{kotler2011single}, one can appreciate that allowing the system to evolve for a time $T$ results in a phase acquired by the qubit given by:
\begin{equation} \label{eq_accumphase_time}
    \Phi = \int g(t) h(t) \mathrm{d}t,
\end{equation}
where we've defined
\begin{equation} \label{eq_hdef}
    h(t) \equiv \text{rect}\left(\frac{t-T/2}{T}\right)\cos\left[\int_0^t \Omega(t') \mathrm{d}t'\right].
\end{equation}
Thus we may regard the accumulated phase as the output of a linear system, with input $g(t)$ filtered by the kernel $h(t)$ \cite{biercuk2011dynamical}.

An especially useful and well-studied scenario is that of dynamical decoupling (DD) in the ``bang-bang'' regime \cite{ViolaPRA1998,ViolaPRL1999,VitaliPRA1999,souza2012robust}, in which $\Omega(t)$ is taken to be a sequence of scaled Dirac delta functions chosen such that the cosine term in Eq. (\ref{eq_hdef}) alternates between $\pm1$. This is physically realized by a train of $\pi$ pulses applied to the qubit in the limit of fast Rabi nutation. In the present study we will constrain ourselves to this template, allotting the freedom to choose the points in time at which $h(t)$ changes sign. The most commonly employed choice has the form:
\begin{equation} \label{eq_freq_filterset}
    h_{j,\phi}(t) = \text{rect}\left(\frac{t-T/2}{T}\right) \text{sgn}\left[\cos\left(2\pi t f_j - \phi\right)\right],
\end{equation}
with modulation frequency $f_j$ and phase $\phi$. This basic DD unit has long been used to prolong qubit coherence by decoupling from low-frequency noise \cite{Falci_PRA_2004,WitzelPRL2007,RyanPRL2010,deLangeScience2010,SagiPRL2010,AlvarezPRA2010,bluhm2011dephasing,NaydenovPRB2011,AjoyPRA2011}. If a signal of interest is also modulated at frequency $f_j$ then lock-in amplification is enabled \cite{taylor2008high,kotler2011single}. Information on the spectrum of $g$ can be recovered via a sequence of measurements with different $f_j$ \cite{almog2011direct,Alvarez_PRL_2011,bylander2011noise,deLangePRL2011,TaminiauPRL2012,bar2012suppression,KotlerPRL2013,RomachPRL2015,MalinowskiPRL2017,HernandezGomezPRB2018}. If the phase of $g$ is unknown then a pair of measurements corresponding to $\phi = 0$ and $\phi = \pi/2$ for each $f_j$ suffices. With this in mind we will drop the $\phi$ subscript for simplicity.

\begin{figure}
    \centering
    \includegraphics{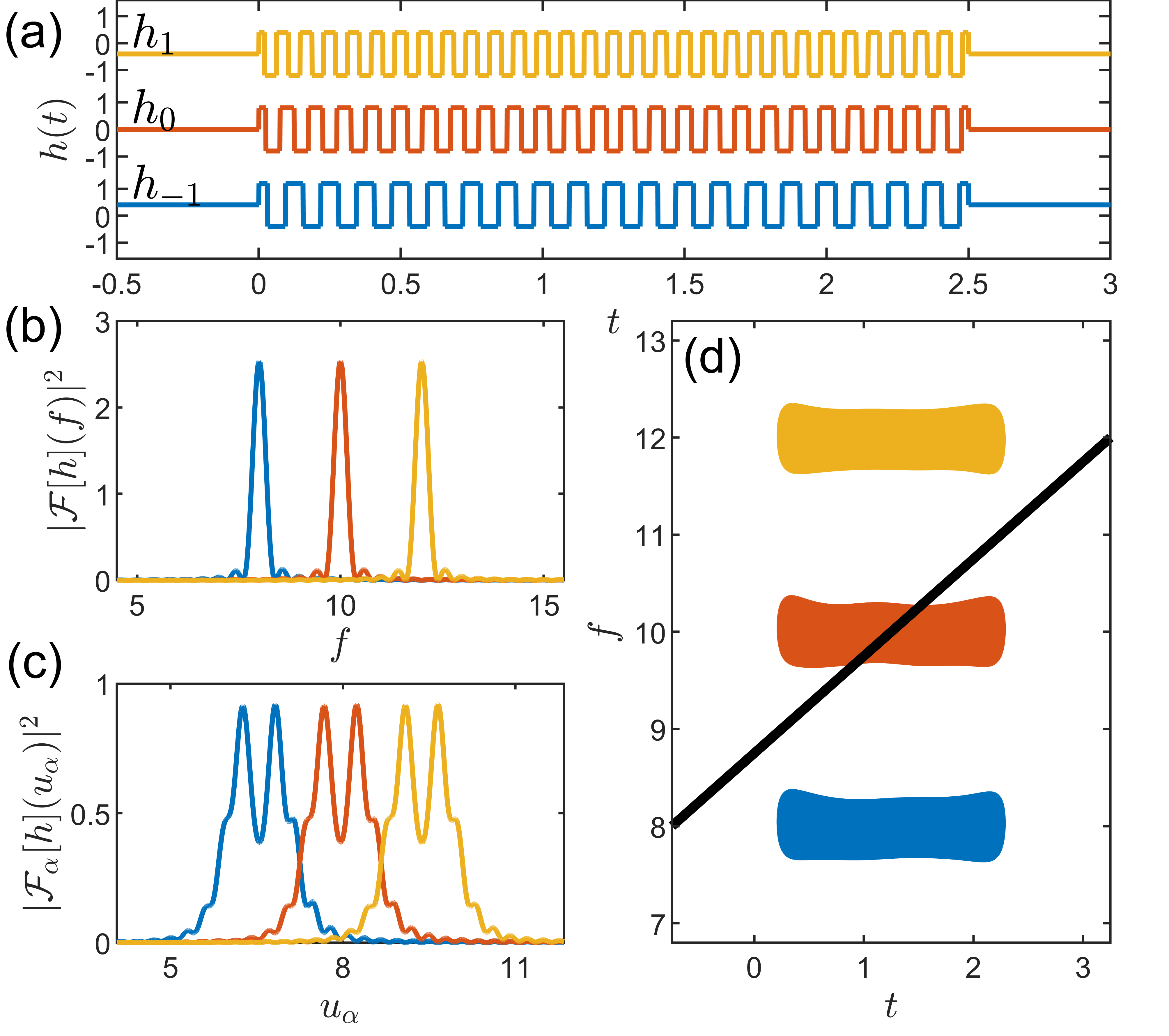}
    \caption{Three examples of ordinary DD filters defined in Eq. (\ref{eq_hdef}), as represented in the (a) time domain, (b), frequency domain, (c) and FRFT domain of order $\alpha = \pi/4$. (d) Filled blobs are bounded by the half-max contours of the Wigner representations of these same three filters. Black line is parallel to the $u_{\alpha=\pi/4}$-axis. Units are arbitrary.}
    \label{fig:fourier filter set}
\end{figure}

Such a sequence of filters is depicted in the time domain in Fig. \ref{fig:fourier filter set}(a). The abrupt sign changes mean that in the frequency domain these idealized filters possess nonzero energy near the odd harmonics of $f_j$. However, if $g$ is sufficiently narrow-band around some frequency $f_0$, and if $f_j$ is not too far from $f_0$, then the phase $\Phi_j$ accumulated under application of the filter $h_j$ is essentially proportional to the Fourier transform of $g$ evaluated at $f_j$ in the limit of large $T$ (Fig. S1 \cite{supp}). More precisely,
\begin{equation}
    \left|\Phi_j\right| \to \frac{4}{\pi}\Big| \mathcal{F}[g](f_j) \Big|
\end{equation}
under such conditions. Equivalently, $\left| \mathcal{F}[h_j](f) \right|$ approaches a Dirac delta centered on $f_j$ in the immediate vicinity of $f_j$ when $T$ is large [Fig. \ref{fig:fourier filter set}(b)]. If $g$ is broad in the frequency domain such that its overlap with the odd harmonics of $f_j$ must be considered, a sequence of measurements followed by inversion of a matrix equation can decompose the spectrum \cite{Alvarez_PRL_2011}.

The fractional Fourier transform (FRFT) is a generalization of the ordinary Fourier transform that has found applications in optics and signal processing \cite{OzaktasFRFT}. We define the FRFT of order $\alpha$ of some $\mathscr{G}(t)$ via:
\begin{eqnarray}
    \mathcal{F}_\alpha[\mathscr{G}](u_\alpha) = &&\sqrt{1-i\cot{(\alpha)}} e^{i \pi \cot{(\alpha)} u_\alpha^2} \times \\ &&\int \mathscr{G}(t)e^{-2\pi i \left( \csc{(\alpha)}u_\alpha t - \frac{\cot{(\alpha)}}{2}t^2 \right)} \mathrm{d}t \nonumber.
\end{eqnarray}
The ordinary Fourier transform is a special case of the FRFT for which $\alpha = \pi/2$. When $\alpha = 0$, the FRFT kernel behaves like a delta function and so leaves the input in the time domain. For other values of $\alpha$ the FRFT effects a transformation of $\mathscr{G}$ into an intermediate time-frequency domain. Square moduli of the $(\alpha=\pi/4)$-order FRFTs of the same filters considered in Fig. \ref{fig:fourier filter set}(a-b) are illustrated in Fig. \ref{fig:fourier filter set}(c). Some nice geometrical properties of the FRFT can be appreciated if we first introduce some concepts from time-frequency analysis \cite{CohenTimeFreq,OzaktasFRFT}. The Wigner distribution representation of $\mathscr{G}$ is defined by:
\begin{equation} \label{eq_Wigner_defn}
    W_\mathscr{G}(t,f) \equiv \int \mathscr{G}(t+t'/2)\mathscr{G}^*(t-t'/2)e^{-2\pi i f t'} \mathrm{d}t'.
\end{equation}
Readers familiar with phase-space approaches to quantum mechanics will recognize Eq. (\ref{eq_Wigner_defn}) from its original context \cite{WignerPR1932}; its use was later adapted for signal processing \cite{ville1948theorie}. The Wigner function represents something like the energy density of $\mathscr{G}$ in the time-frequency plane, with the caveat that the energy near a particular point $(t,f)$ can only be determined to within a window of dimensions limited by the uncertainty principle $\Delta t \Delta f \gtrsim 1$. The integral projection of $W_\mathscr{G}$ onto an axis making an angle $\alpha$ with the time axis yields $|\mathcal{F}_\alpha[\mathscr{G}](u_\alpha)|^2$, the power spectrum in the $\alpha$-order FRFT domain. Figure \ref{fig:fourier filter set}(d) depicts contours of the (smoothed) Wigner representations of the filters described in Fig. \ref{fig:fourier filter set}(a-c) and defined by Eq. (\ref{eq_freq_filterset}). The diagonal black line in Fig. \ref{fig:fourier filter set}(d) is oriented at an angle $\alpha = \pi/4$ from the time axis. From this diagram we can appreciate that the widths of these filters represented in the $u_\alpha$ domain will be intermediate to those in the time and frequency domains. Having introduced the Wigner distribution, we can rewrite the magnitude of the accumulated phase defined in Eq. (\ref{eq_accumphase_time}) as \cite{CohenTimeFreq,OzaktasFRFT}:
\begin{equation} \label{eq_magPhi_wigner}
    |\Phi| = \left( \iint W_g(t,f) W_h(t,f) \, \mathrm{d}t \, \mathrm{d}f \right)^{1/2}, 
\end{equation}
where we've made use of the assumption that $g(t)$ and $h(t)$ are both real. By inspection of Eq. (\ref{eq_magPhi_wigner}), large $|\Phi|$ requires significant overlap of the Wigner functions.
\begin{figure}
    \centering
    \includegraphics{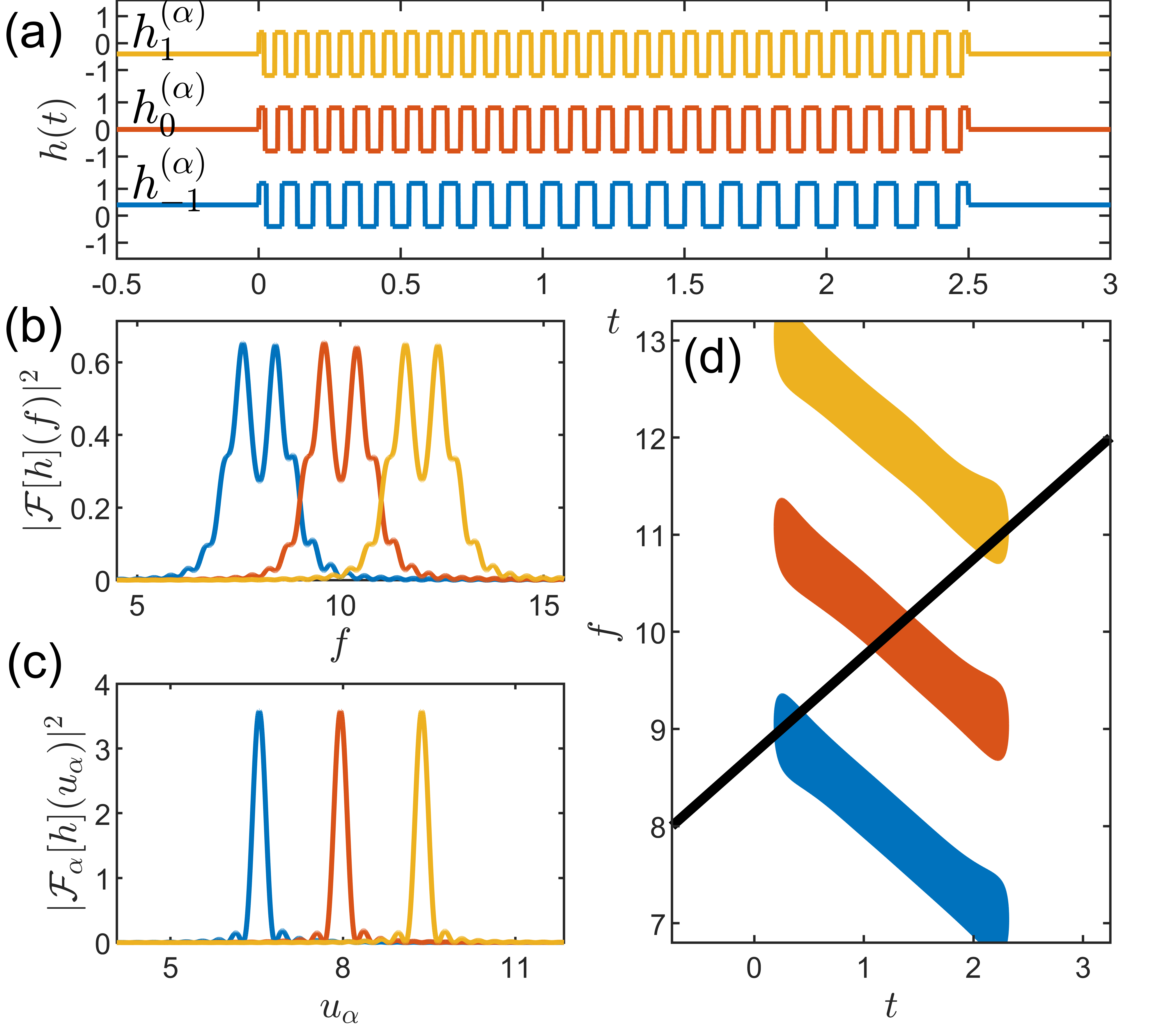}
    \caption{Same explanation as for Fig. \ref{fig:fourier filter set}, except for filters with $\alpha = \pi/4$.}
    \label{fig:FRFT filter set}
\end{figure}

\section{\label{section_resultsanddiscussion}Results and Discussion\\}
Here we introduce a generalized set of linearly chirped DD filters that produces a measurement in the FRFT domain of order $\alpha$. We define:
\begin{eqnarray} \label{eq_FRFT_filterset}
    h_{j,\phi}^{(\alpha)}(t) =&& \text{rect}\left(\frac{t-T/2}{T}\right)\times \\ &&\text{sgn}\left\{\cos\left[2\pi t \left(-\frac{q}{2}t + f_j\right) -\phi\right]\right\} \nonumber,
\end{eqnarray}
where $q \equiv \cot\alpha$. We will henceforth suppress the subscript $\phi$ by the same reasoning as before. Examples of filters of this form are shown in the time domain in Fig. \ref{fig:FRFT filter set}(a). Since the frequency changes over time, the power spectrum in the frequency domain is understandably broadened [Fig. \ref{fig:FRFT filter set}(b)]. By contrast, the power spectrum is narrow in the FRFT domain of order $\alpha = \text{arccot}(q)$ in the vicinity immediately surrounding $u_{\alpha,j} \equiv f_j \sin\alpha$ for sufficiently large $T$ [Fig. \ref{fig:FRFT filter set}c]. Evidently some care must be taken in considering units, as the numerical value of $\alpha$ depends on the units in which we choose to enumerate $t, f$ and $q$. For the illustration in Fig. \ref{fig:FRFT filter set} we've chosen units such that the FRFT order of interest is $\alpha = \pi/4$. Contours of the (smoothed) Wigner representations of these filters are shown in Fig. \ref{fig:FRFT filter set}(d). Similar to the previous case, the abrupt sign changes in the time domain translate to nonzero energy near the odd harmonics of $u_{\alpha,j}$. The peaks centered on odd harmonics become increasingly broad as one moves out from the fundamental as the chirp rate increases in magnitude \cite{supp}. In effect, one must be slightly more careful in establishing conditions under which $\{h_j^{(\alpha)}\}$ produces a measurement of $g$ in the FRFT domain. If $\mathcal{F}_\alpha[g](u_\alpha)$ is sufficiently narrow-band around some $u_{\alpha,0} = f_0 \sin\alpha$, and if $u_{\alpha,j}$ is not too far from $u_{\alpha,0}$, then
\begin{equation}
    \left|\Phi_j^{(\alpha)}\right| \to \frac{4 \sqrt{|\sin\alpha|}}{\pi} \Big| \mathcal{F}_\alpha[g](f_j \sin\alpha) \Big|
\end{equation}
in the limit of large $T$, with the additional caveat that $|f_0-q t|$ should remain large enough that the signal does not intersect the chirped harmonics or the time axis for $t\in[0,T]$ (Fig. S2-S3 \cite{supp}).

\begin{figure}
    \centering
    \includegraphics{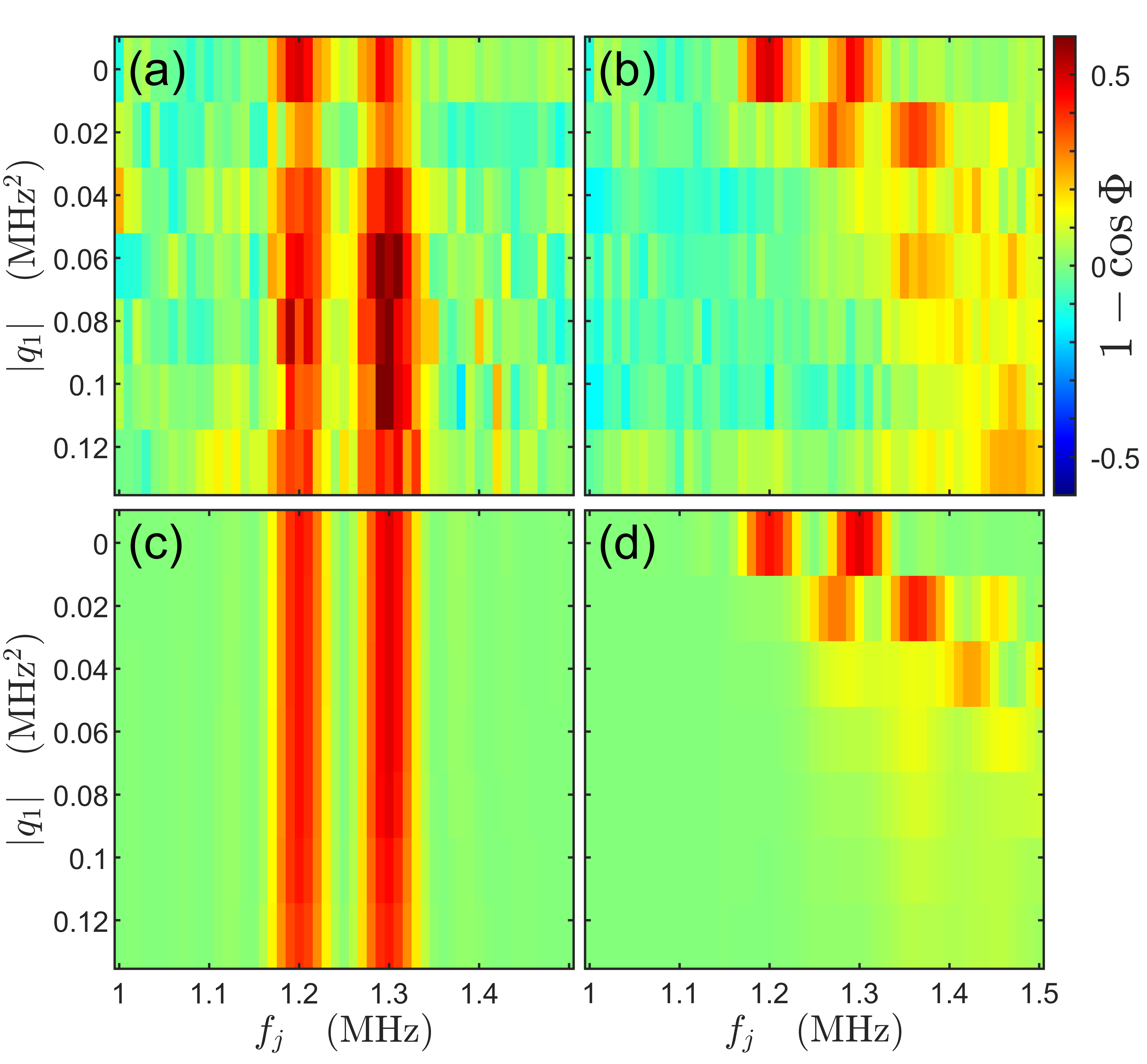}
    \caption{Experimental spectra of AC magnetic field synthesized according to Eq. (\ref{eq_AWG_signal}) for various $q_1$ and $f_1 \in \{1.2, 1.3\}$ MHz, as recorded by an NV ensemble driven with either (a) $q$-matched FRFT sequences, or (b) ordinary unchirped DD sequences. Both show good agreement with calculated spectra in (c) and (d), respectively.}
    \label{fig:experimental spectra comparison}
\end{figure}

With the basic theory now established, we next describe experiments using an NV ensemble that demonstrate an advantage to measurement in the FRFT domain. Signals of the form 
\begin{equation} \label{eq_AWG_signal}
    g(t) = A \cos\left[2\pi t \left(\frac{-q_1}{2}t + f_1\right)\right]\text{rect}\left(\frac{t-T/2}{T}\right),
\end{equation}
(where $T = 9.6 \, \mu$s was fixed, $f_1 \in \{1.2, 1.3\}$ MHz, and $q_1$ was varied over the interval $[-0.125,0]$ MHz$^2$ from experiment to experiment) were synthesized using an arbitrary waveform generator (AWG) and delivered to the NVs via a loop placed near the diamond. Here $A = \gamma_{\text{NV}} B_1$, where $\gamma_\text{NV} = 2\pi\times2.8$ MHz/G is the NV electron's gyromagnetic ratio and $B_1$ is the AC magnetic field amplitude felt by the NVs. Across all experiments, an average field amplitude of $B_1 = 1.38 \, \mu T$ was delivered to the NVs. The sign of $q_1$ specifies up-chirped waveforms according to our definitions. For each pair $(q_1,f_1)$, we performed NV measurements with DD sequences of the form in Eq. (\ref{eq_FRFT_filterset}) with matched chirp rate $q=q_1$, and again with conventional DD sequences (i.e. $q = 0$). These sequences were nested between initialization and readout pulses such that the resulting contrast in NV photoluminescence was proportional to $\cos\Phi$. Background-corrected spectra obtained by averaging together results for the two choices of $f_1$ in post-processing are shown in Fig. \ref{fig:experimental spectra comparison}(a-b). For comparison, spectra predicted from calculations are shown in Fig. \ref{fig:experimental spectra comparison}(c-d). Figure \ref{fig:experimental spectra comparison}(b) and (d) corresponds to measurement with ordinary unchirped filters. As $|q_1|$ increases, these spectra broaden and recede into the noise. By contrast, Fig. \ref{fig:experimental spectra comparison}(a) and (c) corresponds to the case in which the chirp rate of the filter is matched to that of the signal. The two peaks corresponding to $f_1 = 1.2$ and 1.3 MHz are clearly identifiable above the noise and resolvable from one another in each case. Some geometrical intuition can be gleaned upon inspection of the relevant Wigner distributions in the $t$-$f$ plane (Fig. S4).

More quantitative assessments can be made by considering some illustrative estimation and detection tasks. The spectra depicted in Fig. \ref{fig:experimental spectra comparison}(a-b) contain merged data recorded separately with $f_1 = 1.2$ and 1.3 MHz for the sake of visualization. Below we analyze the individual data sets. Moreover, each row in Fig. \ref{fig:experimental spectra comparison}(a-b) depicts the mean of 10 independently measured spectra for both $f_1$ values (i.e. $2\times10 = 20$ in total). To generate statistics we now treat each independent measurement separately. 

\begin{figure}
    \centering
    \includegraphics[width=8.5 cm]{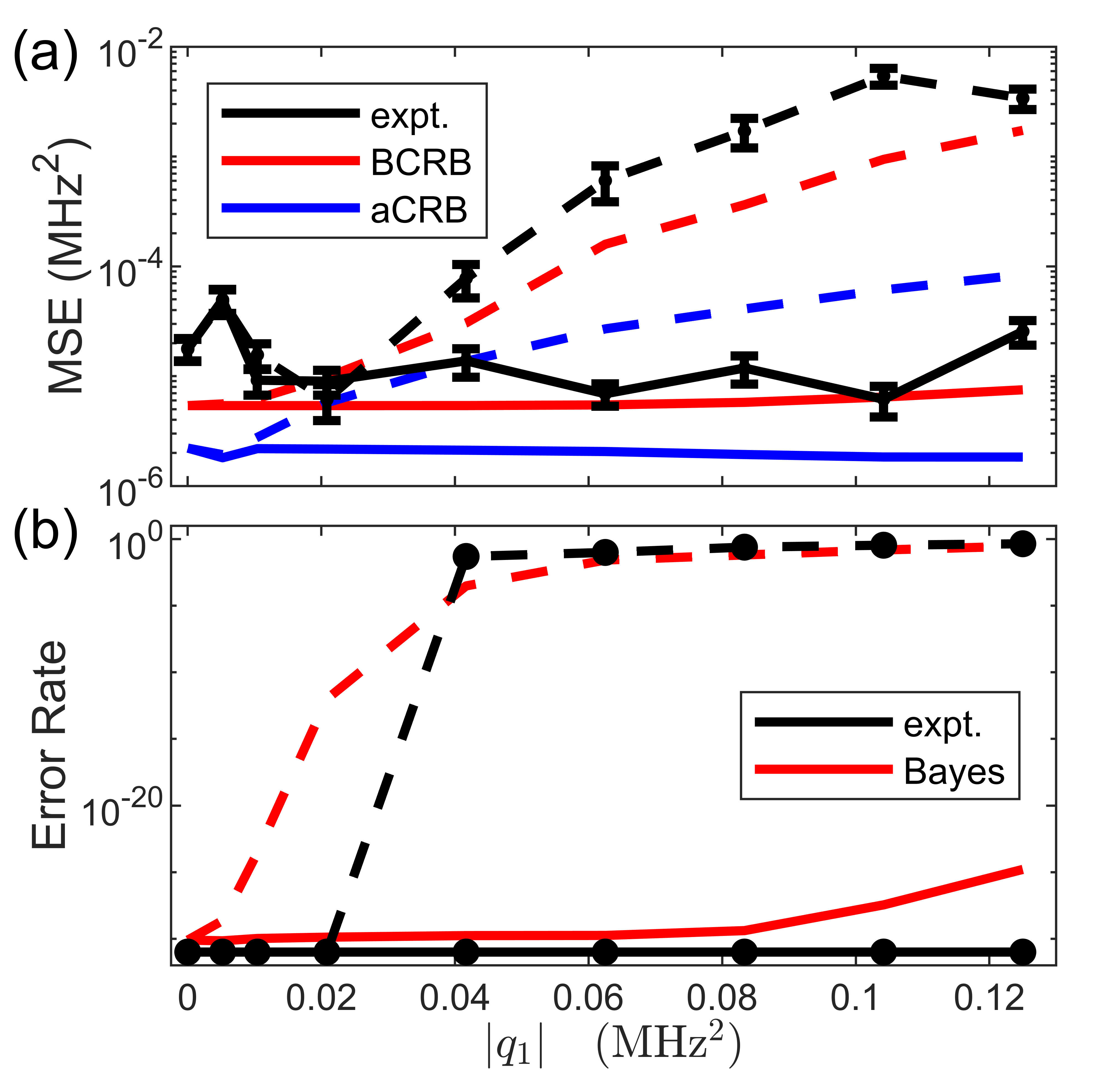}
    \caption{Statistical analysis of experimental data based on $q$-matched FRFT (solid) and unchirped DD (dashed) measurements. (a) Mean-squared error of least-squares estimates of $f_1$ (black). For each $q_1$, 10 measurements were recorded with ground truth $f_1 = 1.2$ MHz and 10 with $f_1 = 1.3$ MHz. Black dots and error bars mark the means and standard deviations, respectively, of 1000 bootstrapped resamples. Red lines depict Bayesian CRBs as described in the text. Blue lines depict the CRBs of the sampling-frequency-adapted measurement described in the text. (b) Experimentally realized MAP error rates (black) for tests of binary hypotheses $f_1 = 1.2$ MHz and $f_1 = 1.3$ MHz. If no errors were recorded across the 20 trials the data point is plotted at $10^{-31}$. Red lines depict the Bayes error rates for this binary hypothesis test.} 
    \label{fig:statistics}
\end{figure}

We acquired least-squares fits for each realization to a model of the form in Eq. (\ref{eq_AWG_signal}), with $A$ and $f_1$ treated as free parameters. Figure \ref{fig:statistics}(a) shows the experimental mean-squared error (MSE) in our estimates of $f_1$ as a function of $q_1$. At large $|q_1|$, the MSEs of the estimates produced from $q$-matched measurements are approximately two orders-of-magnitude smaller than those generated from conventional DD measurements. We next compare these results to limits set by different forms of the Cram\'{e}r-Rao bound (CRB) \cite{CoverInfoTheory,vanTrees2001detection}. We empirically determined that the contrast produced by a particular filter $h_j^{(\alpha)}$ is approximately normally distributed with mean $\cos\Phi_j^{(\alpha)}$ and constant variance of 0.1493. The least-squares estimator of $f_1$ displays significant biases at large $|q_1|$ for the measurement using unchirped filters, likely due to the fact that the signal energy moves toward and through the edge of our sampling range. As the ordinary CRB guarantees a lower bound to the variance of \textit{unbiased} estimators, it is not necessarily appropriate to compare to this entity. A Bayesian CRB can instead be computed that bounds the MSE for any estimator, biased or unbiased \cite{vanTrees2001detection}. Taking a Gaussian prior for $f_1$ and numerically extremizing the BCRB vs. the variance of the prior produces the values plotted in red in Fig. \ref{fig:statistics}(a). These bounds recapitulate an improvement in MSE of roughly two orders-of-magnitude at high absolute chirp rates. The BCRB for the unchirped filters could be made tighter by treating $A$ as a nuisance parameter with a prior distribution of its own in our calculation, but the dashed red line serves as a useful aid to the eye nonetheless.

An adaptive measurement scheme in which the sample frequencies are redrawn to match the distribution of signal energy in a given domain would improve sensitivity in both cases. We computed the CRB for estimation of $f_1$ according to a scheme in which a fixed number of sample frequencies were evenly redistributed over a range containing 95\% of the signal energy, effectively making the sample range for matched-chirp measurements narrower and that for unchirped filters broader. The results are plotted in blue in Fig. \ref{fig:statistics}(a). Compared in this way, the gap in performance at high $|q_1|$ is somewhat smaller, but still close to two orders-of-magnitude.

We next analyze our experimental data from the vantage of a binary hypothesis test in which a signal of the form in Eq. (\ref{eq_AWG_signal}) is assumed, and given an observation realized with one set of filters or the other, the observer is to determine whether the signal was generated with parameter $f_1 = 1.2$ or 1.3 MHz. We implemented a maximum \textit{a posteriori} (MAP) hypothesis test to make this decision given our empirically-determined Gaussian noise model. Results of this test are shown in Fig. \ref{fig:statistics}(b) in black. Since only $n = 20$ samples were included at each $q_1$, it is impossible to realize an experimental error rate on the open interval $(0,0.05)$. In cases where no errors were recorded among the 20 trials, the data has been plotted at $10^{-31}$ in Fig. \ref{fig:statistics}(b) for the sake of visualization. No errors were recorded in discriminating the two models when sensing was performed in the correct FRFT domain. In the frequency domain case no errors were recorded for $q_1$ between 0.02 and 0.04 MHz$^2$, after which errors became increasingly frequent. To contextualize these results derived from limited statistics we computed the Bayes error, i.e. the minimum probability of error for the test (red lines), which for our noise model has a simple analytical form in terms of the complementary error function \cite{vanTrees2001detection,supp}.

To this point in our discussion we have considered only deterministic signals, but extension to stochastic $g$ is straightforward. In that case the accumulated phase $\Phi$ defined by Eq. (\ref{eq_accumphase_time}) is itself a random variable. It can easily be shown that \cite{supp}:
\begin{equation}
    \left\langle \Phi^2 \right\rangle = \iint W_g(t,f) W_h(t,f) \, \mathrm{d}t \, \mathrm{d}f, 
\end{equation}
where for stochastic, real-valued $g$ the Wigner function is defined \cite{CohenTimeFreq,OzaktasFRFT}:
\begin{equation}
    W_g(t,f) \equiv \int \left\langle g\left(t+t'/2\right)g\left(t-t'/2\right) \right\rangle e^{-2\pi i f t'} \mathrm{d}t'.
\end{equation}
Note that $W_g$ is independent of $t$ if $g$ is wide-sense stationary. Therefore we only expect sensing of stochastic signals in the FRFT domain to be possibly of interest for non-stationary processes.
\section{\label{Conclusion}Conclusion\\}
We have presented a set of filters and an accompanying theoretical framework to facilitate quantum sensing in the fractional Fourier domain of order $\alpha$. This generalizes conventional approaches to sensing in the time and frequency domains. Our filters are distinct from previously reported aperiodic sequences \cite{KernPRL2005,KhodjastehPRL2005,ViolaPRL2005,DharPRL2006,UhrigPRL2007,GordonPRL2008,UysPRL2009,UhrigPRL2009,PanPRA2010,PasiniPRA2010,ClausenPRL2010,WestPRL2010,KhodjastehPRA2011,HayesPRA2011,GracePRA2012}, most of which were specifically engineered to decouple from noise of certain characteristics, though some have been applied \cite{soare2014experimental,XuPRB2016} or proposed \cite{HallPRB2010,zhao2011atomic,CasanovaPRA2015,NorrisPRA2018} for sensing as well. Using an NV ensemble, we experimentally implemented measurements in the fractional Fourier domain and demonstrated significant metrological advantages for some proof-of-principle inference tasks. These filters may find utility in quantum adaptations of classical applications that employ chirped waveforms, such as radar \cite{klauder1960theory} and spread-spectrum communication protocols \cite{ReyndersIEEE}. 

For NVs specifically, these pulse sequences could engender new capabilities in nanoscale nuclear magnetic resonance (NMR) studies \cite{bucher2019quantum,allert2022advances,aslam2023quantum}. The magnetic field at the NV due to the precession of nuclei confined to a nanoscale volume can often be treated as a mean-zero classical stochastic process with correlation functions proportional to those of the total nuclear spin quantum operator \cite{meriles2010imaging,PhamPRB2016,HernandezGomezPRB2018}. If the Zeeman term dominates the nuclear Hamiltonian and the bias field, $B_0$, is swept linearly in synchrony with the NV measurement protocol (with the frequency of the NV drive modulated to compensate for this sweep), then the Wigner distribution corresponding to the stochastic field generated by the precessing nuclei will exhibit a linear chirp. A procedure commonly employed in NV-detected NMR for confirming the presence of nuclei of a particular gyromagnetic ratio is to slowly and discretely vary $B_0$ between independent measurements, then perform a linear regression on the extracted Larmor frequencies of each trial \cite{mamin2013nanoscale,staudacher2013nuclear,loretz2014nanoscale,devience2015nanoscale}. Instead, by sweeping $B_0$ quickly during each measurement and sensing in the FRFT domain, this procedure could be performed with lock-in amplification. If $B_0$ is sufficiently small such that the nuclear Zeeman term competes with internuclear couplings, then the sweep in $B_0$ might accompany a level crossing or avoided crossing, and the shape of the stochastic signal in the time-frequency plane would depend on the adiabaticity of this process. Perhaps this effect could be mined for chemical and physical information.

\begin{acknowledgments}
This work was supported in part by the Arnold and Mabel Beckman Foundation, and by a Spectroscopy Society of Pittsburgh Starter Grant Award. L.P.A. acknowledges additional support from the Thor R. Rubin Fellowship, the Lester E. and Kathleen A. Coleman Fellowship, and the Victor E. Buhrke Graduate Fellowship.

S.H. and M.P.B. conceived of and performed the experiments. M.P.B. did the theory. D. D. wrote control software and contributed to the building of the experimental setup. L. P. A. prepared the diamond substrate and contributed to the building of the experimental setup. R. S. contributed to early versions of the basic experiment.
\end{acknowledgments}

\bibliography{mybib}
\end{document}